# Unmasking Inequity: Socio-Economic Determinants and Gender Disparities in Maharashtra and India's Health Outcomes – Insights from NFHS-5

**Sharmishtha Raghuvanshi**

Mumbai School of Economics and Public Policy,

University of Mumbai, Mumbai

mishuraghuvanshi27@gmail.com

**Supriya Sanjay Nikam**

Research Fellow,

Mumbai School of Economics and Public Policy,

University of Mumbai, Mumbai

supriyanikam2002@gmail.com

**Manisha Karne**

Professor,

Mumbai School of Economics and Public Policy,

University of Mumbai, Mumbai

manisha.karne@gmail.com

**Satyanarayan Kishan Kothe**

Professor,

Mumbai School of Economics and Public Policy,

University of Mumbai, Mumbai

kothesk@gmail.com

**Unmasking Inequity: Socio-Economic Determinants and Gender Disparities in Maharashtra and India's Health Outcomes – Insights from NFHS-5**

## Abstract

This research examines the persistent challenge of health inequalities in India, departing from the conventional focus on aggregate improvements in mortality rates. While India has achieved progress in overall health indicators since independence, the distribution of health outcomes remains uneven, a fact starkly highlighted by the COVID-19 pandemic. This study investigates the socio-economic determinants of health disparities using the National Family and Health Survey (NFHS)-5 data from 2019-20, focusing on both national and state-level analyses, specifically for Maharashtra.

Employing a health economics framework, the analysis delves into individual-level data, population shares, self-reported morbidity prevalence, and treatment patterns across diverse socio-economic groups. Regression analyses, stratified by gender, are conducted to quantify the impact of socio-economic factors on reported morbidity. Furthermore, a Fairlie decomposition, an extension of the Oaxaca decomposition, is utilised to dissect the gender gap in morbidity, assessing the extent to which observed differences are attributable to explanatory variables.

The findings reveal a significant burden of self-reported morbidity, with approximately one in nine individuals in India and one in eight in Maharashtra reporting morbidity. Notably, women exhibit nearly double the morbidity rate compared to men. The decomposition analysis identifies key drivers of gender disparities. In India, marital status exacerbates these differences, while insurance coverage, caste, urban residence, and wealth mitigate them. In Maharashtra, urban residence and marital status widen the gap, whereas religion, caste, and insurance coverage narrow it. This research underscores the importance of targeted policy interventions to address the complex interplay of socio-economic factors driving health inequalities in India.

## Keywords

Self-reported morbidities, socio-economic factors, health inequalities, Fairlie decomposition, logit, logistic regression, Blinder-Oaxaca decomposition, health equality, health in Maharashtra, Health in India, Morbidity, Gender inequality in health

## Introduction

Health inequalities in India persist as a critical public health challenge, despite aggregate improvements in mortality and health indicators since independence. While national averages indicate progress, the distribution of health outcomes remains strikingly uneven across socio-economic strata—a disparity that the COVID-19 pandemic has exacerbated. Existing literature has predominantly focused on mortality trends or macro-level health metrics, often overlooking the nuanced dynamics of self-reported morbidity and its socio-economic determinants. This study addresses this gap by leveraging the National Family and Health Survey (NFHS)-5 (2019–21) to dissect the structural drivers of health inequalities, with a dual focus on national trends and Maharashtra, a state emblematic of India's economic diversity and regional heterogeneity.

The measurement of health inequalities in India has long relied on data from sources such as the Census, National Sample Survey (NSS), and NFHS, which provide critical insights into mortality, morbidity, and healthcare utilisation (Saikia & Kulkarni, 2016). Nevertheless, self-reported morbidity—a crucial indicator of population health—remains inadequately explored due to fragmented data availability. While the Census offers decadal snapshots, the NSS and NFHS provide more frequent but ailment-specific morbidity data, limiting longitudinal analysis (Akhtar et al., 2020; Anushree & Madheswaran, 2019). Nevertheless, these datasets have catalysed pivotal research on socio-economic determinants of health, revealing persistent disparities linked to caste, wealth, insurance coverage, and rural-urban divides (Paul & Singh, 2017; Prinja et al., 2012; Bhojani et al., 2013). For instance, lower-income and marginalised caste groups disproportionately bear the burden of untreated morbidity, exacerbated by out-of-pocket healthcare expenditures that consume over 60% of household health spending (Oxfam, 2021; Akhtar et al., 2020).

State-level analyses further underscore the limitations of national aggregates. Studies in Punjab, Karnataka, and Kerala reveal divergent patterns in morbidity, healthcare access, and insurance penetration, emphasising the need for granular, region-specific policymaking (Bango & Ghosh, 2023; Anushree & Madheswaran, 2019; Prinja et al., 2012). Maharashtra, a state marked by stark socio-economic disparities alongside rapid urbanisation and industrial expansion, reflects India's broader health inequities in a concentrated form. Despite the state's significant economic stature, its health outcomes and care accessibility remain inadequately examined through rigorous empirical and policy lenses. The non-communicable diseases (NCDs) like diabetes and hypertension have been on the rise in recent years in India. This shift is deepening health inequities, as under-resourced health systems and socioeconomic disparities disproportionately hinder vulnerable populations' access to preventive care and disease management. This burden disproportionately impacts socioeconomically marginalised groups, who face systemic barriers including fragmented healthcare infrastructure, prohibitive out-of-pocket expenditures, and geographic or cultural impediments to service utilisation. The intersection of these factors underscores a critical misalignment between the growing NCD burden and the health system's capacity to deliver equitable, affordable care. Addressing this disparity necessitates targeted fiscal policies, strengthened primary care networks, and interventions that directly confront the social determinants perpetuating these vulnerabilities. The COVID-19 pandemic exposed these fissures, as comorbidities amplified mortality risks and strained an overburdened public health system (Gopalan & Misra, 2020). Concurrently, disparities in healthcare access worsened: marginalised groups faced barriers to testing and treatment, while insurance coverage—concentrated among wealthier households—failed to mitigate financial shocks (Oxfam, 2021; Mukherjee & Krishanu, 2008). These dynamics underscore the urgency of disentangling the socio-economic and gendered drivers of health disparities.

Prior studies have employed concentration indices, regression decompositions, and Blinder-Oaxaca models to quantify health inequalities (Brinda et al., 2016; Jain et al., 2013). However, the Fairlie decomposition—a nonlinear extension of Oaxaca's method—remains underutilised in analysing gendered health disparities, particularly in the Indian context. This approach is critical for disentangling how observed variables (e.g., marital status, insurance) differentially shape morbidity reporting between men and women. Furthermore, while existing research highlights wealth and caste as key determinants, interactions between multiple disadvantages (e.g., low-income + Scheduled Caste status) remain inadequately explored (Mohindra et al., 2006; Roy et al., 2004).

The discussion above raises a question - to understand and analyse the nature and extent of health inequalities prevailing in India and Maharashtra across various socio-economic indicators and to what extent do the inequality outcomes pertain to differences based on gender groups and how much of these differences can be explained by socio-economic indicators?

This research advances the literature in three key ways. First, it shifts the focus from mortality to self-reported morbidity—a sensitive indicator of healthcare access and health-seeking behaviour—using the latest NFHS-5 data. Second, it employs a health economics framework to conduct parallel national and state-level (Maharashtra) analyses, addressing the limitations of extrapolating national findings to subnational contexts (Jha et al., 2022). Third, it applies Fairlie decomposition to quantify gendered disparities, identifying policy-actionable factors that exacerbate or mitigate gaps. By integrating individual-level socio-economic variables (e.g., insurance, urban residence, caste) and contextualising findings within India's post-COVID health landscape, this study provides a roadmap for targeted, equity-oriented interventions.

The analysis reveals that approximately one in nine Indians and one in eight Maharashtrians report morbidity, with women's rates doubling those of men. Regression and decomposition models pinpoint marital status, insurance coverage, and caste as pivotal determinants, with distinct patterns at national and state levels. These insights refine our understanding of health inequities and inform strategies to strengthen India's healthcare financing and delivery systems amid ongoing epidemiological and demographic transitions.

## Data and Methodology

National Family Health Survey 5 (NFHS-5) data collected for 2019-21 has been leveraged for the present study. The data was collected using multiple stratified sampling done over two different phases. The data consists of 636,699 households, consisting of 2,843,917 members. Individual-level health issues related data is collected for 724,115 women aged 15-49 and 101,839 men aged 15-54. Further data related to Tuberculosis and Disabilities has been mapped from the household-level data for the same set of individuals. The health status of a population is reflected in the levels of morbidity and treatment-seeking behaviour of its members. Based on the availability of data for this paper self-reported morbidity consists of the following health problems diabetes, hypertension, chronic respiratory disease including asthma, thyroid disorder, heart disease, cancer, chronic kidney disorder, tuberculosis and the disabilities categorised as hearing, speech, visual, mental, locomotor and others. The prevalence of co-morbidities has not been considered in this paper. If an individual suffers from more than one form of illness or disability or a combination of the two, they are still assigned a morbidity value of 1, indicating that "yes" they are ailing from a condition or disability and only those that do not suffer from any condition or disability are assigned the morbidity value of 0.

The variables considered in the data are categorical and have been divided into groups. Based on gender the data is classified into male and female. Based on age the data has been divided into 5-year age groups starting from 15 years to 49 years for women and 15 years to 54 years for men. Based on the type of residence the groups are classified into rural and urban. Based on educational attainment the groups are those with no education and primary, secondary and higher level of education. Based on religion the data is divided into Hindus, Muslims Christians and all other religions in the "Others" category. Based on the wealth index five groups have been created poorest, poorer, middle, richer and richest. Insurance coverage divides data into two groups, those covered by insurance and those not covered by insurance. Based on castes the division is between those belonging to Schedule Tribe /Schedule Caste (ST/SC), Other Backward class (OBC) and forward castes. Based on literacy level, the groups are divided into those who cannot read at all, those who can read parts, those who can read whole sentences and the residual categories where data cards were not available for the language spoken by the individual and those visually impaired. Based on marital status, the data is divided into those who are married, those who were never in union of any form, widowed, divorced and those who are separated and or no longer live together. Based on household size individuals are classified based on the members in the households to which they belong, those with less than or equal to 5 members, between 6 to 10 members, 11 to 15 members and above 16 members. These classifications are consistent for data on a country level for India and a state level for Maharashtra. At a country level, the data for states is grouped into standard regions central, north, east, north-east, south and west.

**Measures of Health Inequality**

The following analysis has been performed to understand the demographic characteristics of the individuals in the data, their health status and treatment patterns:

1. Regression Analysis: Decomposition of the measures is performed using regression analysis of the morbidity variable which is binary (1 for if the individual has reported morbidity and 0 if the individual does not have any morbid condition or the same has not been reported.) against the explanatory variables to find their coefficients then calculate the measures for the health variable and its determinants. Our dependent variable is binary which is why the Logit model is used to study the correlation between various socio-economic factors and the prevalence of self-reported morbidity at a country level for India and state level for Maharashtra. The independent variables are categorical in nature and the base variable has been decided based on which had the lowest level of collinearity with other sets of variables using the inflation factor. The coefficient as well as the odds ratio has been considered for the interpretation of the results of our regression model. The below equations present the logit model with the first term being the intercept term followed by variable terms where j = 9 in our model and u represents the error term.

$$Logit(Y) = \ln\left(\frac{p}{1-p}\right) = \beta_0 + \Sigma \sum_{j=1}^{J} \beta_j \,.\, x_{j,k} + u_k$$

2. Decomposition Analysis: We have performed decomposition for two groups based on gender, Group 1 representing Men and Group 2 representing Women. The independent variables are the same that have been considered for regression analysis. The analysis was performed using the "Fairlie" code in STATA/MP 13.0. For non-linear models, they suggest the extension of the Blinder-Oaxaca model given by Fairlie should be referred to. Fairlie (2005) provided an extension of the Blinder-Oaxaca decomposition technique for logit and probit models. He acknowledges that the Blinder-Oaxaca model is extremely useful for studying gender gaps in outcomes for linear models and also provides a comparison of Blinder-Oaxaca results and his own extension model results. The Blinder-Oaxaca decomposition may not perform well in cases where differences are located in the tails of the distribution or differences in independent variables are very large. Sharma et al. (2021) used Fairlie decomposition analysis to study income-based inequalities in the immunisation of children. Their results show that maternal education and place of residence had a positive contribution, and caste and age of the child had a negative contribution to the inequality gap for full immunisation between lower and higher-income groups.

A measure of inequality that can be decomposed into different groups or regions. This equation is an extension of the Blinder-Oaxaca decomposition method for non-linear models like in our case (logit model) given by Fairlie (2005).

$$\triangle \bar{Y} = [\frac{1}{N^1}\sum_{i=1}^{N^1} F(\beta^1 X_i^1) - \frac{1}{N^2}\sum_{i=1}^{N^2} F(\beta^1 X_i^2)] + [\frac{1}{N^2}\sum_{i=1}^{N^2} F(\beta^1 X_i^2) - \frac{1}{N^2}\sum_{i=1}^{N^2} F(\beta^2 X_i^2)]$$

Where N is the sample size, with two groups 1 and 2 and Y represents the difference in "mean predicted probability of outcome" between the two groups. The decomposition method provides data on how much and in which direction (widening the differences or narrowing them) each of the variables contribute to the differences in probability.

## Empirical Evidence: Regression Analysis

Due to differences in the categories of variables and sample size of data collected for men and women a separate model has been run for men and women in India as well as Maharashtra. The variables into consideration are age groups, place of residence, educational attainment, religion, caste, household size, wealth index, insurance coverage and marital status. The base variables for these categorical variables were selected in a manner to reduce multicollinearity in the model. Variance Inflation Factors (VIF) were calculated and all the individual variables had a VIF value of a maximum value of 2.33 and a Mean VIF for the model of a maximum of 1.62. A link test (it works on the premise that if our model is specified correctly, then if we were to regress the outcome variable on the prediction and the prediction squared, the prediction squared would have no explanatory power) was performed for all models and the results showed that the prediction variable was significant at 99 percent significance and prediction squared variable were not significant at 90 percent significance level.

According to the results of logistic regression and odds ratios obtained are presented in Table 4 and Table 5 for India and Maharashtra respectively; the following inferences can be made. In India, females were more likely to report morbidities than males. Morbidities were more likely among the population residing in urban areas as compared to rural areas in India with an odds ratio of greater than 1 amongst men as well as women. Women in urban areas were more likely to report morbidity than men in urban areas. With an increase in age, the odds ratio also increased for both men and women indicating that individuals from older age groups reported higher morbidity.

**Table 1: Results of regression analysis for men and women reporting morbidities in India across socio-economic variables**

| Attributes | Women (n = 724115) | | | | Men (n = 101839) | | | |
|---|---|---|---|---|---|---|---|---|
| | **Coef.** | **Robust SE** | **Odds Ratio** | **Robust SE** | **Coef.** | **Robust SE** | **Odds Ratio** | **Robust SE** |
| Rural | 0# | | 1# | | 0# | | 1# | |
| Urban | 0.1244415*** | 0.0094294 | 1.132516*** | 0.010679 | 0.0098166 | 0.032826 | 1.009865 | 0.03315 |
| Age 15-19 | 0# | | 1# | | 0# | | 1# | |
| Age 20-24 | 0.1972605*** | 0.0186194 | 1.218061*** | 0.02268 | 0.3430215*** | 0.077906 | 1.409199*** | 0.109786 |
| Age 25-29 | 0.4623972*** | 0.0202338 | 1.587876*** | 0.032129 | 0.5902815*** | 0.080259 | 1.804496*** | 0.144827 |
| Age 30-34 | 0.7560003*** | 0.0209092 | 2.129741*** | 0.044531 | 0.899009*** | 0.082394 | 2.457167*** | 0.202456 |
| Age 35-39 | 1.001932*** | 0.020825 | 2.723539*** | 0.056718 | 1.127834*** | 0.083733 | 3.088957*** | 0.258647 |
| Age 40-44 | 1.243828*** | 0.0210781 | 3.468865*** | 0.073117 | 1.548523*** | 0.082957 | 4.704515*** | 0.390271 |
| Age 45-49 | 1.44431*** | 0.0210385 | 4.238926*** | 0.089181 | 1.784688*** | 0.082464 | 5.957719*** | 0.491298 |
| Age 50-54 | 0# | | 1# | | 2.10275*** | 0.082856 | 8.18866*** | 0.678477 |
| No education | -0.2573534*** | 0.0104844 | 0.773095*** | 0.008106 | -0.1506286*** | 0.04185 | 0.860167*** | 0.035998 |
| Primary | -0.0380052*** | 0.0116626 | 0.962708*** | 0.011228 | -0.0162434 | 0.040223 | 0.983888 | 0.039575 |
| Secondary | 0# | | 1# | | 0# | | 1# | |
| Higher | -0.1405266*** | 0.0126764 | 0.868901*** | 0.011015 | -0.0044044 | 0.038374 | 0.995605 | 0.038206 |
| Hindu | 0# | | 1# | | 0# | | 1# | |
| Muslim | 0.2003768*** | 0.0112179 | 1.221863*** | 0.013707 | 0.114018*** | 0.041378 | 1.120772*** | 0.046376 |
| Christian | 0.2201928*** | 0.0144415 | 1.246317*** | 0.017999 | 0.198419*** | 0.051266 | 1.219473*** | 0.062518 |
| Other religions | 0.2322519*** | 0.0162872 | 1.261437*** | 0.020545 | 0.1820982*** | 0.056924 | 1.199732*** | 0.068293 |
| Members 1-5 | 0# | | 1# | | 0# | | 1# | |
| Members 6-10 | 0.0402068*** | 0.0079194 | 1.041026*** | 0.008244 | -0.1641056*** | 0.028999 | 0.848652*** | 0.02461 |
| Members 11-15 | 0.1748157*** | 0.0209342 | 1.191027*** | 0.024933 | 0.0180439 | 0.078242 | 1.018208 | 0.079667 |
| Members 16 and above | 0.0538623 | 0.0524797 | 1.055339 | 0.055384 | 0.4353902** | 0.188742 | 1.545566** | 0.291713 |
| Poorest | -0.0430375*** | 0.0119489 | 0.957876*** | 0.011446 | -0.1768101*** | 0.043947 | 0.837939*** | 0.036825 |
| Poorer | 0# | | 1# | | 0# | | 1# | |
| Middle | 0.0638987*** | 0.0113439 | 1.065984*** | 0.012092 | 0.0213537 | 0.040025 | 1.021583 | 0.040889 |
| Richer | 0.0887251*** | 0.0119448 | 1.09278*** | 0.013053 | 0.0951966** | 0.041618 | 1.099875** | 0.045775 |
| Richest | 0.0613304*** | 0.0137181 | 1.06325*** | 0.014586 | 0.1224861** | 0.047371 | 1.130303** | 0.053544 |
| Not covered by health insurance | 0# | | 1# | | 0# | | 1# | |
| Covered by health insurance | 0.0861807 *** | 0.0077467 | 1.090003 *** | 0.008444 | 0.1541386 *** | 0.026391 | 1.166653 *** | 0.030789 |
| Never in union | -0.1281484 *** | 0.0156695 | 0.879723 *** | 0.013785 | -0.2111966 *** | 0.052683 | 0.809615 *** | 0.042653 |
| Married | 0# | | 1# | | 0# | | 1# | |
| Widowed | 0.0720444 *** | 0.0182165 | 1.074703 *** | 0.019577 | -0.1757134 | 0.126934 | 0.838858 | 0.106479 |
| Divorced/ No longer living together/separated | 0.2656667 *** | 0.0293741 | 1.3043*** | 0.038313 | 0.1807225 | 0.120558 | 1.198083 | 0.144439 |
| ST/SC | -0.0048312 | 0.0093319 | 0.99518*** | 0.009287 | -0.0589431 * | 0.032973 | 0.942761 * | 0.031086 |
| OBC | 0# | | 1# | | 0# | | 1# | |
| Forward Castes | 0.1501797 *** | 0.0094551 | 1.162043 *** | 0.010987 | 0.1379237 *** | 0.033397 | 1.147888 *** | 0.038337 |
| Constant | -2.809795*** | 0.0207404 | 0.060217 *** | 0.001249 | -3.79266*** | 0.08323 | 0.022536 *** | 0.001876 |

**Authors' Calculation: Detailed results of regression analysis**
**#Base Variable**
*****Significant at 99% Confidence level. p-value < 0.01**
****Significant at 95% Confidence level. p-value < 0.05**
***Significant at 90% Confidence level. p-value < 0.1**

**Table 2: Results of regression analysis for men and women reporting morbidities in Maharashtra across socio-economic variables**

| Attributes | Women (n = 33755) | | | | Men (n = 5482) | | | |
|---|---|---|---|---|---|---|---|---|
| | Coef. | Robust SE | Odds Ratio | Robust SE | Coef. | Robust SE | Odds Ratio | Robust SE |
| Rural | 0# | | 1# | | 0# | | 1# | |
| Urban | 0.0110817 | 0.0411894 | 1.011143 | 0.041648 | 0.0779442 | 0.140421 | 1.081062 | 0.151804 |
| Age 15-19 | 0# | | 1# | | 0# | | 1# | |
| Age 20-24 | 0.1418633* | 0.0804366 | 1.152419* | 0.092697 | 0.9146134*** | 0.322485 | 2.49581*** | 0.804863 |
| Age 25-29 | 0.3899741*** | 0.0863929 | 1.476942*** | 0.127597 | 0.8191985** | 0.339255 | 2.268681** | 0.769661 |
| Age 30-34 | 0.6393498*** | 0.0877313 | 1.895248*** | 0.166273 | 1.247758*** | 0.337546 | 3.482526*** | 1.175511 |
| Age 35-39 | 0.8475514*** | 0.0886314 | 2.333925*** | 0.206859 | 1.467758*** | 0.347945 | 4.339497*** | 1.509905 |
| Age 40-44 | 1.048411*** | 0.0896897 | 2.853114*** | 0.255895 | 1.887119*** | 0.34511 | 6.600324*** | 2.277837 |
| Age 45-49 | 1.212863*** | 0.0897012 | 3.363098*** | 0.301674 | 2.100116*** | 0.342108 | 8.167121*** | 2.794036 |
| Age 50-54 | 0# | | 1# | | 2.23145*** | 0.345775 | 9.313361*** | 3.220324 |
| No education | -0.0818086 | 0.0524421 | 0.9214483 | 0.048323 | -0.5571914** | 0.238235 | 0.5728156** | 0.136464 |
| Primary | -0.0212582 | 0.0516583 | 0.9789662 | 0.050572 | -0.1518717 | 0.171288 | 0.8590985 | 0.147153 |
| Secondary | 0# | | 1# | | 0# | | 1# | |
| Higher | -0.1555568*** | 0.0556794 | 0.8559384*** | 0.047658 | -0.0882174 | 0.156441 | 0.9155618 | 0.143232 |
| Hindu | 0# | | 1# | | 0# | | 1# | |
| Muslim | 0.2154686*** | 0.0545919 | 1.240443*** | 0.067718 | -0.5800391** | 0.232447 | 0.5598765** | 0.130142 |
| Christian | 0.3422594* | 0.2046335 | 1.408126* | 0.28815 | -0.5954023 | 0.718165 | 0.5513407 | 0.395954 |
| Other religions | 0.2146037*** | 0.0622072 | 1.239371*** | 0.077098 | -0.1623664 | 0.214687 | 0.8501297 | 0.182512 |
| Members 1-5 | 0# | | 1# | | 0# | | 1# | |
| Members 6-10 | 0.2056624*** | 0.035107 | 1.228338*** | 0.043123 | -0.1286034 | 0.123257 | 0.8793226 | 0.108383 |
| Members 11-15 | 0.7763196*** | 0.0746124 | 2.173458*** | 0.162167 | 0.4261255 | 0.282267 | 1.531313 | 0.432239 |
| Members 16 and above | 0.6988275*** | 0.187309 | 2.011393*** | 0.376752 | 0## | | 1## | |
| Poorest | -0.0087606 | 0.0510009 | 0.9912776 | 0.050556 | 0.2523457 | 0.16494 | 1.287041 | 0.212284 |
| Poorer | -0.0411764 | 0.0480239 | 0.9596598 | 0.046087 | -0.00985 | 0.164481 | 0.9901984 | 0.162869 |
| Middle | 0# | | 1# | | 0# | | 1# | |
| Richer | 0.0013915 | 0.0501548 | 1.001392 | 0.050225 | -0.017488 | 0.176509 | 0.982664 | 0.173449 |
| Richest | -0.1255703** | 0.0631612 | 0.8819937*** | 0.055708 | 0.2537771 | 0.207369 | 1.288884 | 0.267275 |
| Not covered by health insurance | 0# | | 1# | | 0# | | 1# | |
| Covered by health insurance | 0.2003719*** | 0.0414289 | 1.221857*** | 0.05062 | 0.0358877 | 0.127642 | 1.036539 | 0.132306 |
| Never in union | 0.103858 | 0.070301 | 1.109443 | 0.077995 | -0.1014279 | 0.191684 | 0.9035463 | 0.173196 |
| Married | 0# | | 1# | | 0# | | 1# | |
| Widowed | -0.0732621 | 0.0796124 | 0.9293572 | 0.073988 | 1.43233*** | 0.457899 | 4.188446*** | 1.917886 |
| Divorced/ No longer living together/separated | 0.1310445 | 0.1200204 | 1.140018 | 0.136826 | -1.168114 | 1.017302 | 0.3109528 | 0.316333 |
| ST/SC | -0.0983061** | 0.0445092 | 0.9063714** | 0.040342 | -0.0740438 | 0.144858 | 0.928631 | 0.13452 |
| OBC | 0.0342236 | 0.0395078 | 1.034816 | 0.040883 | -0.0345502 | 0.127581 | 0.9660399 | 0.123248 |
| Forward Castes | 0# | | 1# | | 0# | | 1# | |
| Constant | -2.560447*** | 0.0885496 | 0.0772702*** | 0.006842 | -3.87528*** | 0.353121 | 0.020749*** | 0.007327 |

**Authors' Calculation: Detailed results of regression analysis**
**#Base Variable**
*****Significant at 99% Confidence level. p-value < 0.01**
****Significant at 95% Confidence level. p-value < 0.05**
***Significant at 90% Confidence level. p-value < 0.1**

An increase in education level as compared to people with no education did contribute positively to reporting self-morbidities for both men and women. The contribution to self-reported morbidities decreased as the level of education increased for men while for women they increased till secondary education and the odds ratio fell for women with higher educational attainment. For women, those from religions other than Hindu, Muslim and Christian have the highest odds ratio indicating a higher probability as compared to others to report morbidity. Among men, Christians indicated the highest probability to report morbidity, and both men and women reported the lowest probability to report morbidity. A similar result is found in the work of Paul and Singh (2017).

An increase in the wealth index led to an increase in self-reported morbidity. With those from the poorest wealth category contributing negatively to reporting morbidity. An interesting inference from the results is that women with the poorest and middle wealth index were more likely to report morbidity than men but as the wealth index increased, men from the richer and richest wealth index category were more likely to report morbidity than women. Both men and women who have never been in a union were least likely to report morbidity with an odds ratio of less than 1 as compared to those married and or divorced/separated. Widowed men were also less likely to report morbidity, while widowed women contributed positively to reporting morbidity. A similar result was found in the work of Paul and Singh (2017). Amongst both men and women, those who were divorced/separated were more likely to report morbidity. Women hailing from households with 11-15 members were more likely to report morbidity as compared to other women from other household size categories. Men belonging to households with more than 16 members were likely to report morbidity as compared to other women from other household size categories.

The model for men in Maharashtra, omitted data from 15 observations which consist of those men who belong to households having more than or equal to 16 members, since none of the individuals in this category reported any ailments or disabilities. In Maharashtra as well, females were more likely to report morbidities than males. Similarly, morbidities were more likely among the population residing in urban areas as compared to rural areas, women from urban areas were more likely to report morbidity than men from urban areas. Similar to India, in Maharashtra as well, with an increase in age group the men as well as women were more likely to report morbidity.

For the Education attainment variable, amongst the four categories defined, no defined trend was seen to establish a relationship between the impact of increase in level of educational attainment and reporting morbidity. Morbidities were more likely women who have never been in a union or divorced/separated as compared to men who have never been in a union or divorced/ separated in Maharashtra. Widowed Men were more likely to report morbidity as compared to Widowed Men.

For Maharashtra, the men from the richest and poorest wealth index groups in terms of more likely to report morbidities as compared to men from other wealth index groups. Women from richer wealth groups were more likely to report morbidity while those from the richest wealth group were least likely to report morbidity. As the wealth index increased the contribution to self-reported morbidities increased. As the members in household increased, the contribution of an individual to reporting morbidity also increased.

Women from the Christian religion were more likely to report morbidity as compared to any other religion and those from the Hindu religion were least likely to report morbidity, while among men those from the Hindu religion were more likely to report morbidity as compared to any other religion. Both men and women belonging to households with 11-15 members were most likely to report morbidity. While for women, those belonging to household size of 1-5 members were least likely to report morbidity and for men, those belonging to household size of 6-10 members were least likely to report morbidity.

Individuals from the ST/SC class were less likely to report morbidities than those from OBC in India and well as Maharashtra. Individuals from forward castes were more likely to report morbidities as compared to backward classes in India. In Maharashtra, OBC women were more likely to report morbidity as compared to women from forward classes (similar results found in the work of (Mohindra et al., 2006) while men from forward castes were more likely to report morbidity as compared to OBC men. As supported by prior evidence for both men and women who are covered by insurance were more likely to report morbidities as compared to those not covered by insurance. This has been supported by similar evidence found by Akhtar et al., 2020, Oxfam 2021 and Brinda et al (2015). In India, women who are covered by insurance were less likely to report morbidity compared to men who were covered by insurance. While in Maharashtra, women who were covered by insurance were more likely to report morbidity than men who were covered by Insurance.

**Table 3: ROC (receiver operating character) graph derived for logit regression model in STATA**

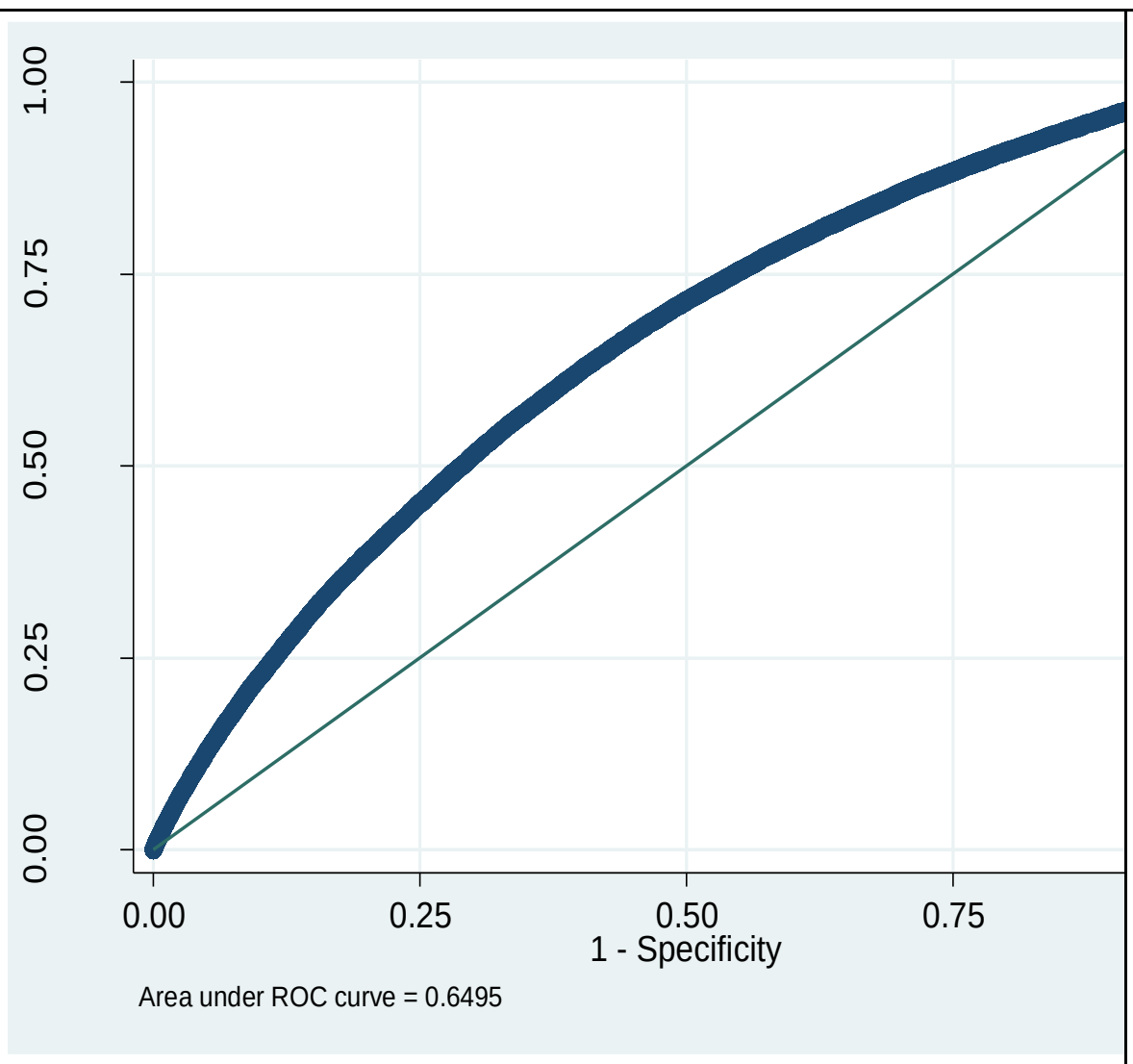

**Logit Regression for Women in India**

number of observations = 724115

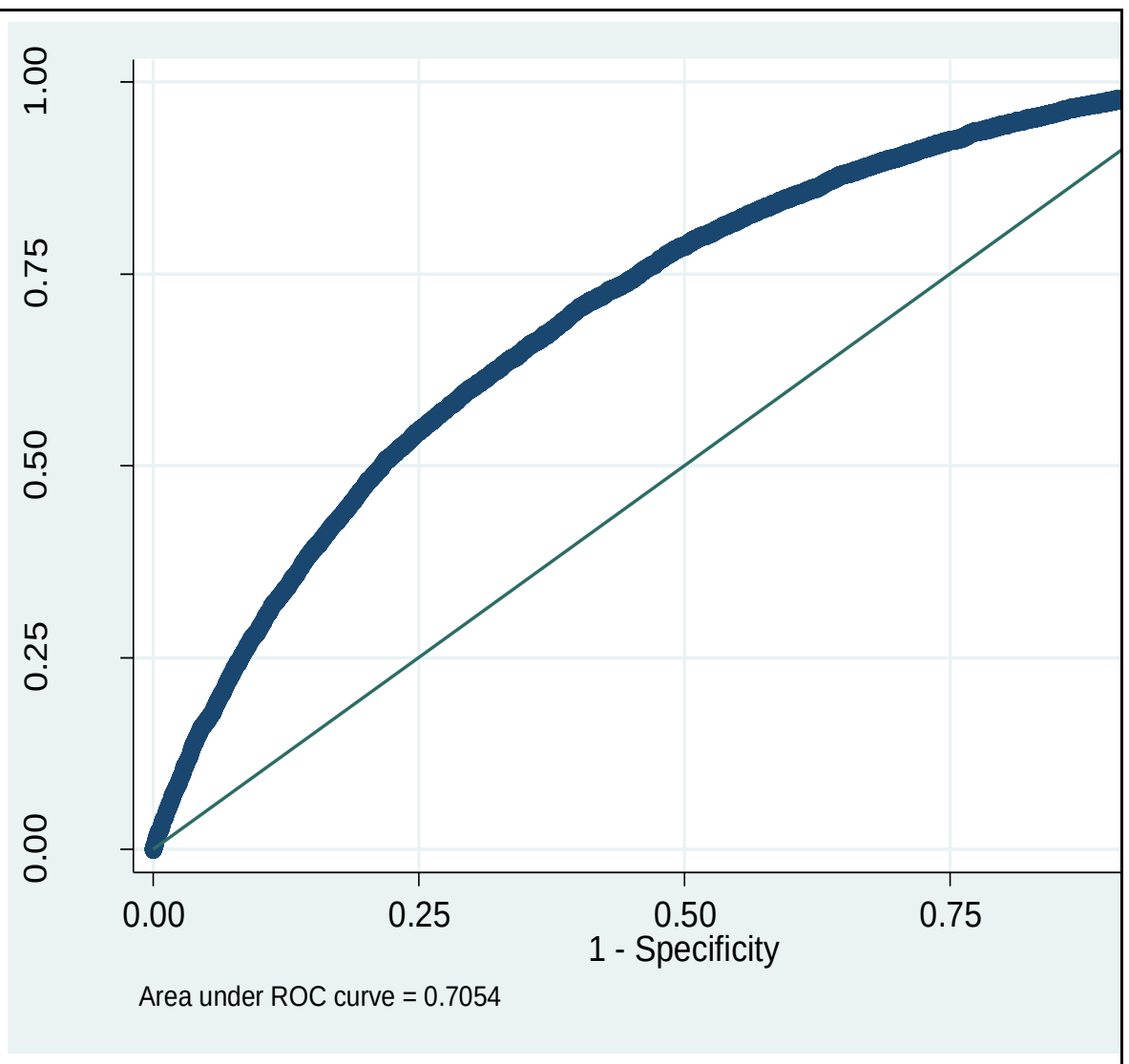

**Logit Regression for Men in India**

number of observations =101839

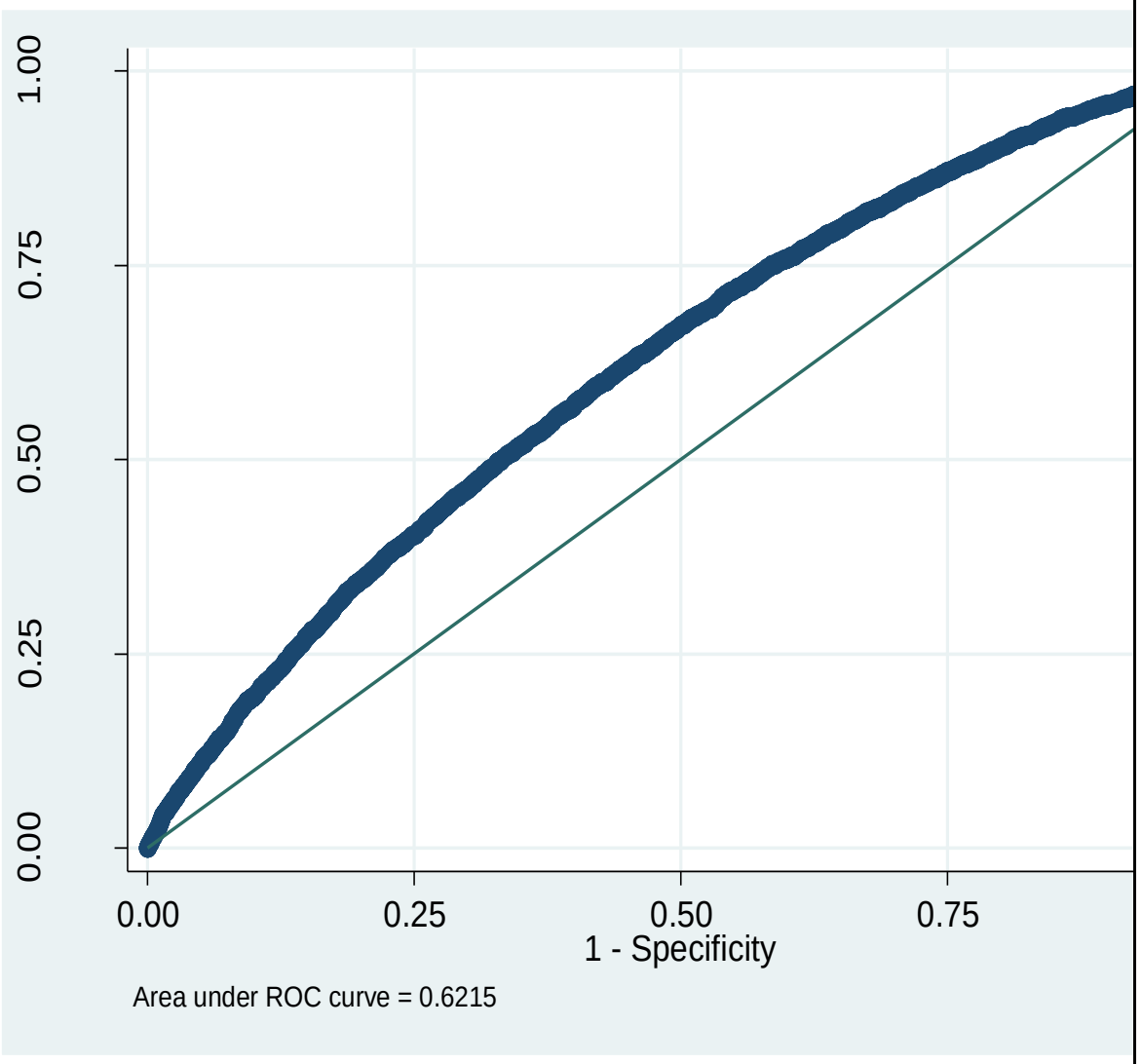

**Logit Regression for Women in Maharashtra**

number of observations = 33755

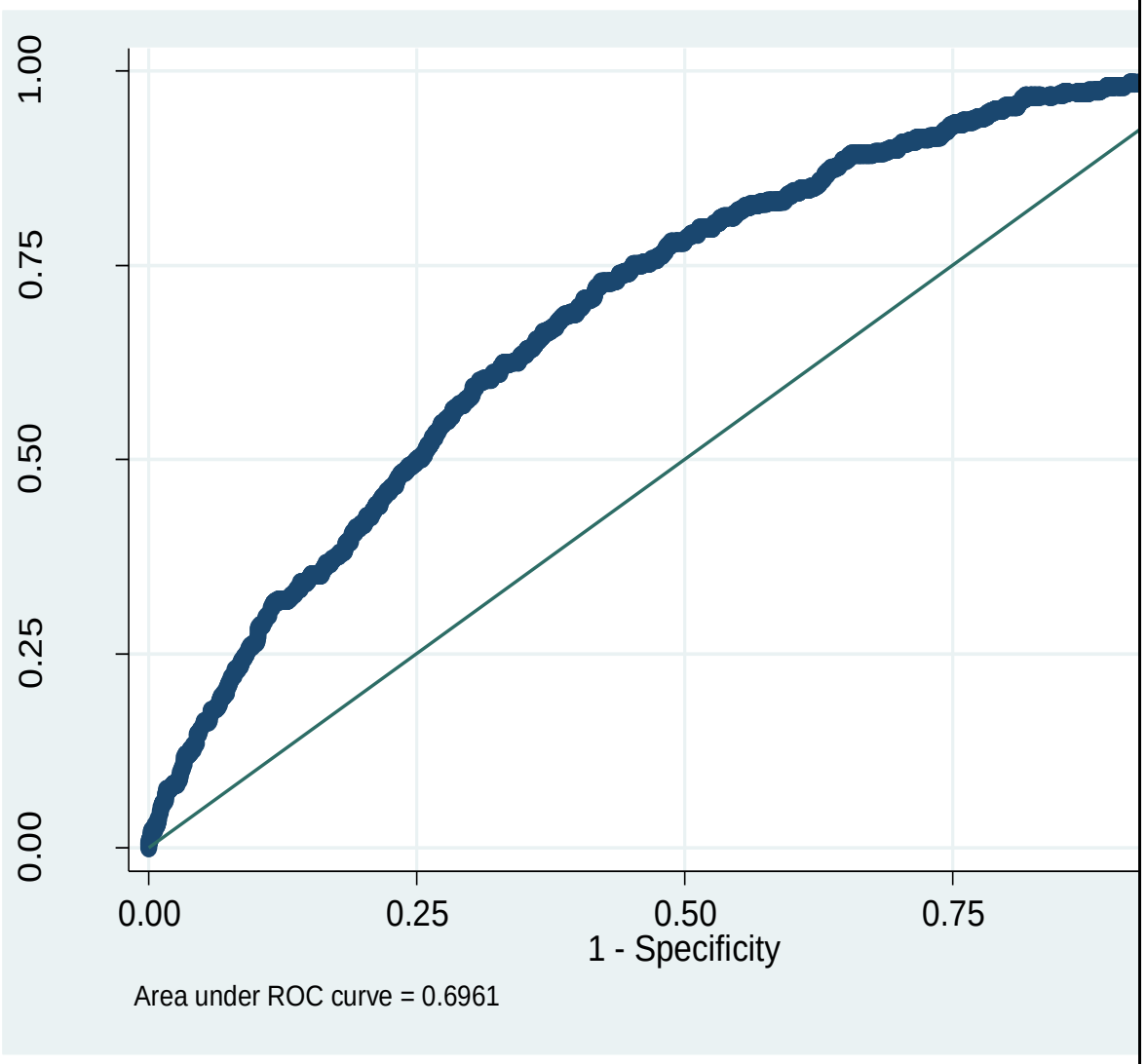

**Logit Regression for Men in Maharashtra**

number of observations = 5482

**Source: Authors' Calculation**

The ROC (receiver operating character) curve for our regression models was derived, it is a graph of sensitivity on the Y-axis versus one minus specificity on the X-axis and calculates the area under it.

Sensitivity refers to the probability that the model predicts a positive outcome when indeed the outcome is positive, while specificity is the probability that the model predicts a negative outcome when indeed the outcome is negative. (Bobbitt, 2020). A value of 0.5 for AUC (area under the curve) indicates that the ROC curve will fall on the diagonal and indicates that the diagnostic test has no discriminatory ability (Mandrekar, 2010) since all our AUCs are greater than 0.5, our model has discriminatory power and the regression results are significant, hence the null hypothesis, that morbidity outcomes are not different across socio-economic groups, can be rejected.

**Decomposition Analysis:**

Since our outcome variable is binary and model being non-linear in nature the Fairlie method is used to perform decomposition analysis for India and Maharashtra. The method decomposes data into two groups and calculates the conditional probability and differences between the groups. It goes a step further and analyses to what extent these differences can be explained by the independent variables. The independent variables are the same as is used for regression analysis, the base variables for categorical variables have also been kept the same except in the case of wealth index where medium wealth index is taken as the base variable and caste where the base variable is taken as OBC for India and Maharashtra. This was done to ensure a model with minimum mean VIF is selected. The mean VIF for the model for India and well as Maharashtra was 1.58 and the individual variable VIF of maximum 2.26.

**Table 4: Summary results of decomposition for men and women reporting morbidity**

| **Details** | **India** | **Maharashtra** |
| --- | --- | --- |
| Number of observations | 8,25,954 | 39,252 |
| No. of observations - Male | 1,01,839 | 5,497 |
| No. of observations - Female | 7,24,115 | 33,755 |
| Pr(Morbidity!=0\|G=Male) | 0.0659178 | 0.07294888 |
| Pr(Morbidity!=0\|G=Female) | 0.1235260 | 0.14214190 |
| Difference | -0.0576082 | -0.06919302 |
| Total explained | 0.0036942 | -0.00199674 |
| % explained | -6.4126 | 2.88575 |

**Source: Authors' Calculation: Summary Results from decomposition**

The above details in Table 7, show a summary of running decomposition analysis for our model using Fairlie code in STATA for non-linear decomposition analysis grouping by Women. The first three rows provide data related observations. Pr(Morbidity!=0|G=Male) gives the probability that morbidity is 1(the individual suffers from an illness or disability) given that they are a Male. Similarly, Pr(Morbidity!=0|G=Female) gives the probability given they are a Female. The same values can be seen in our Demographic analysis as well in the "Morbidity prevalence per 100" column for gender-wise bifurcation. The last three rows calculate the difference in the two probabilities calculated and "Total Explained" provides data on to what extent those differences are explained by the socio-economic indicators we have considered in our model, details for the same are provided in the following table. The explained portion of 0.0036942 and -0.00199674 has been further broken down into the following variables:

**Table 5: Results of decomposition analysis between differences in Men and women reporting morbidities across various socio-economic factors**

| Attributes | India | | | Maharashtra | | |
|---|---|---|---|---|---|---|
| | Coef. | Robust SE | % contribution | Coef. | Robust SE | % contribution |
| Rural | 0# | | | 0# | | |
| Urban | 0.0002918*** | 0.0000210 | -0.5065 | -0.0000543 | 0.0001046 | 0.0785 |
| Age 15-19 | 0# | | | 0# | | |
| Age 20-24 | -0.0004266*** | 0.0000568 | 0.7405 | -0.0001361 | 0.0001317 | 0.1967 |
| Age 25-29 | -0.001724*** | 0.0001051 | 2.9926 | -0.001005*** | 0.0003512 | 1.4525 |
| Age 30-34 | -0.0030098*** | 0.0001208 | 5.2246 | -0.0020057*** | 0.0004460 | 2.8987 |
| Age 35-39 | -0.0042281*** | 0.0001313 | 7.3394 | -0.0032828*** | 0.0005624 | 4.7444 |
| Age 40-44 | -0.0041637*** | 0.0001008 | 7.2276 | -0.0032921*** | 0.0004108 | 4.7578 |
| Age 45-49 | 0.0009967*** | 0.0001085 | -1.7301 | -0.0027786*** | 0.0002852 | 4.0157 |
| Age 50-54 | 0.0147197*** | 0.0006184 | -25.5514 | 0.0100622*** | 0.0025041 | -14.5422 |
| No education | 0.0026857*** | 0.0001501 | -4.6620 | 0.0004465 | 0.0005307 | -0.6453 |
| Primary | 0.0000002 | 0.0000013 | -0.0003 | 0.0000063 | 0.0000847 | -0.0091 |
| Secondary | 0# | | | 0# | | |
| Higher | -0.0004857*** | 0.0000429 | 0.8431 | -0.0008893*** | 0.0002918 | 1.2852 |
| Hindu | 0# | | | 0# | | |
| Muslim | 0.0000966*** | 0.0000083 | -0.1677 | 0.0001717** | 0.0000733 | -0.2481 |
| Christian | -0.0000182** | 0.0000076 | 0.0316 | 0.0000527 | 0.0000493 | -0.0762 |
| Other religions | 0.000163*** | 0.0000119 | -0.2829 | 0.0000430 | 0.0000379 | -0.0621 |
| Members 1-5 | 0# | | | 0# | | |
| Members 6-10 | -0.00000693*** | 0.0000030 | 0.0120 | 0.0000340 | 0.0000419 | -0.0491 |
| Members 11-15 | 0.0000427*** | 0.0000063 | -0.0741 | 0.0007236*** | 0.0001110 | -1.0458 |
| Members 16 and above | -0.0000057 | 0.0000041 | 0.0098 | -0.0000768* | 0.0000449 | 0.1110 |
| Poorest | 0.000203*** | 0.0000186 | -0.3524 | -0.0000006 | 0.0000336 | 0.0009 |
| Poorer | 0.00006*** | 0.0000103 | -0.1042 | -0.0000432 | 0.0000525 | 0.0624 |
| Middle | 0# | | | 0# | | |
| Richer | 0.0000132*** | 0.0000050 | -0.0229 | -0.0000098 | 0.0000455 | 0.0142 |
| Richest | 0.0000001 | 0.0000031 | -0.0002 | 0.0000942 | 0.0001027 | -0.1361 |
| Not covered by health insurance | 0# | | | 0# | | |
| Covered by health insurance | 0.0004192*** | 0.0000390 | -0.7277 | 0.0011524*** | 0.0002844 | -1.6655 |
| Never in union | -0.0016118*** | 0.0000927 | 2.7979 | -0.0013373 | 0.0008425 | 1.9327 |
| Married | 0# | | | 0# | | |
| Widowed | -0.0004066*** | 0.0000617 | 0.7058 | -0.0000732 | 0.0003704 | 0.1058 |
| Divorced/ No longer living together/separated | -0.000054*** | 0.0000066 | 0.0937 | -0.0001453 | 0.0001332 | 0.2100 |
| ST/SC | 0.0000156 | 0.0000148 | -0.0271 | 0.0003469** | 0.0001338 | -0.5014 |
| OBC | 0# | | | 0# | | |
| Forward Castes | 0.000115*** | 0.0000150 | -0.1996 | 0.0000379 | 0.0000680 | -0.0548 |

**Source: Authors' Calculation: Detailed results of decomposition**
**#Base Variable**
*****Significant at 99 Confidence level. p-value < 0.01**
****Significant at 95 Confidence level. p-value < 0.05**
***Significant at 90 Confidence level. p-value < 0.1**

In the given case, our difference term is negative for India and Maharashtra. The percentage contribution is calculated using the difference value as a denominator, hence a negative contribution in this case will be a result of a positive coefficient and similarly a positive contribution will be a result of a negative coefficient. A positive contribution presents that a particular variable has increased the gap and a negative contribution presents that a variable has reduced the gap.

As per the decomposition analysis presented in Table 8, for India, urban region, older age groups of age groups 45-49 and 50-54, no and primary educational attainment, Muslim and other religions, households with 11-15 members, wealth index (all groups), coverage by health insurance and caste (all groups) have negative contribution stating that they reduce the difference in probability of men and women reporting morbidity. While younger age groups 20-24, 25-29,30-34, 35-39 and 40-44, Christian religion, higher educational attainment, households with 6-10 and 16 and above members, and marital status (all groups) have a positive contribution indicating that they cause an increase in difference in probability of men and women reporting morbidity. The highest positive contribution was by middle age groups of 35-39 and 40-44 years, followed by the younger age groups of 30-34 and 25-29 years and never in the union variable. The highest negative contribution was by age group 50-54, this is because data for the age group is collected only for men, followed by no educational attainment, age group 45-49, insurance coverage and urban region.

For Maharashtra, the oldest age groups of age group 50-54, no and primary educational attainment, religion (all groups), households with 6-10 and 11-15 members, richest wealth index, coverage by health insurance and caste (all groups) have negative contribution stating that they reduce the difference in probability of men and women reporting morbidity. In the urban region, all age groups except 50-54, age groups 20-24, 25-29,30-34, 35-39 and 40-44, 45-49, higher educational attainment, households 16 or above members, poorest, poorer and richer wealth groups, and marital status (all groups) have a positive contribution indicating that they cause an increase in the difference in probability of men and women reporting morbidity. The highest positive contribution was by age groups of 40-44 years and 35-39, 45-49 and 30-34 years and never in the union variable. The highest negative contribution was by age group 50-54, again similar to India because data for the age group is collected only for men, followed by insurance coverage, households with 11-15 members, no educational attainment, and ST/SC.

**Conclusion**

Self-reported morbidity at national and state level has been sufficiently examined by the inequalities in self-reported morbidity caused socio-economic differences (Akhtar et al., 2020, Anushree and Madheswaran ,2019, Paul and Singh, 2017, Prinja et al. 2012, Mohindra et al., 2006), and gender gap analysis (Sinha et al.,2022, Mishra et al., 2012, Balagopal, 2009) this paper has attempted to fill the research gap by providing detailed decomposition the socio-economic inequalities in self-reported morbidity at the state level for Maharashtra and all India level. It also dealt with decomposing inequalities to reveal its determinants for the population. To summarize our findings, our results showed that women report higher morbidity as compared to men in India as well as Maharashtra. Variable-wise trends were seen in these factors like men were more likely to report morbidity in rural areas and women were more likely to report morbidity in urban areas. In India, the self-reported morbidity rate was 11.62 individuals per 100 and in Maharashtra, 13.25 per 100 individuals reported morbidity, higher as compared to India. The highest prevalence was of Diabetes in India and Hypertension in Maharashtra while the lowest prevalence was of Tuberculosis in both. The highest treatment rate was for Tuberculosis and the lowest was for Cancer in India as well as Maharashtra. Individuals covered by insurance had a higher treatment rate across all morbidities and reported more conditions as well adding to the existing evidence that insurance is an inflating factor for self-reported morbidity (Oxfam, 2021, Akhtar et al., 2020, and Brinda et al.,2015).

The level of insurance coverage is quite low with only 16.61 percent and 32.58 percent of individuals in Maharashtra and India respectively being covered by health insurance.

In India, states in the Southern region reported the highest morbidity as well as highest treatment rate across ailments while states in the Central region reported the lowest morbidity. Lowest treatment rates were reported in states from the North-Eastern region. The gap in rural-urban morbidity reporting was higher in India as compared to Maharashtra, the morbidity rate in Maharashtra had only a marginal gap of 0.23 per 100 but in India as well as Maharashtra the treatment rate across ailments was higher in Urban areas compared to Rural areas. This could be due to higher levels of health awareness and better access to healthcare in urban areas.

Mortality rates have shown a decreasing trend (Tumbe, 2020) (except in 2020-21 which can be attributed to the pandemic) and there has been an increase in life expectancy over the last few decades (O'Neill, 2019) but the gap caused due inequalities in morbidity levels has not reduced. One of the reasons for this is most research and analysis that includes Jain, Singh and Pathak (2013), Joe, Mishra and Navaneetham (2008), Kumar (2022), Prinja, Kanavos, and Kumar (2012) and Tumbe (2020) are performed on either on infant and maternal mortality rates or both and hence the policymaking is also more focused on reducing the maternal and infant mortality rates and in general on the reduction of death rates, hence less focus is given to morbidity levels (Anushree and Madheswaran, 2019). It has also been observed that regions with lower mortality rates have higher reported morbidity rates (Johansson, 1991), this could be attributed as a reason for not seeing a substantial reduction in morbidity levels even after the government has marginally increased spending in areas of healthcare as a percentage of GDP in the last few years (Minhas, A, 2024 and The World Bank, 2023).

The results of the decomposition of differences within the group calculation of percent of their contribution can act as a great policy measure since more focus can be paid to factors that increase the inequalities and the factors that reduce the inequalities can be leveraged. In India, marital status contributed to increasing differences between morbidity reported by men and women, and insurance coverage, caste, urban region and wealth index reduced the differences. In Maharashtra, urban region and marital status widened the differences between morbidity reported by men and women and religion, caste and insurance coverage narrowed this gap. The policy implication of these results is not that simple, for example, no educational attainment is one of the factors in India with negative contribution, policy measures cannot be taken to promote no education solely because it reduces inequality. At the same time, there are factors for which action can be taken, like insurance coverage does have a negative contribution, so policy measures can be taken to increase insurance coverage to help reduce inequalities.

Policy measures like financial assistance directly to backward classes, increasing access to healthcare services by reducing price barriers and setting up local community groups have been suggested to be beneficial in reducing social health inequalities (Mishra et al., 2021 and Anushree and Madheshwaran, 2019). Existing policy measures like Reproductive, Maternal, Newborn, Child, Adolescent Health and Nutrition (RMNCAH+N) Strategy under National Health Mission (NHM) should focus more on adolescent Health and Nutrition If proper amount and direction of focus is given to each of the illnesses India can be successful to reducing them one by one if not completely curbing them, similar to reductions that have been achieved by India for Tuberculosis and Polio. An extremely successful awareness campaign and vaccination drive was carried out for both these illnesses, which has now made our country polio-free and has helped achieve the lowest prevalence rates for Tuberculosis with the highest treatment rates which can be seen in the demographic analysis section of this study as well.

The level of media exposure plays a very important role in changing morbidity reporting and treatment behaviour of individuals. Policy measures can be used by increasing awareness of the illness which firstly increases the level of morbidities reported and corrects for underreporting of morbidities and then through appropriate awareness regarding treatment measures, increases the treatment rate. (Mishra et al., 2012 and Roy et al., 2004)

Lessons can be taken from countries like Australia and Finland, which have more equity in health, these are measures that require a long-term change and require careful steps to implementation in the way decisions related to healthcare are made. Finland has a highly decentralized health provision system where while policy measures at the centre are taken related to grant subsidies, municipalities are being given more autonomy regarding the provision of healthcare services, which helps in targeting inequalities at the smallest geographical level (Koivusalo, 1999). Australia has also focused on developing a more targeted healthcare system, by ensuring Combinations of universal and targeted approaches to ensure that benefits provided at a universal level flow down to a grass-root level where the inequalities exist and target to curbing them. (Fisher et al., 2022). This can be achieved by implementing policy measures for incentives and increasing investment in the healthcare sector to run illness-specific and municipal-level targeted programs for awareness and treatment subsidies.